\documentclass[%
 reprint,
 amsmath,amssymb,
 aps,
 superscriptaddress, 
]{revtex4-1}

\usepackage{graphicx}
\usepackage{dcolumn}
\usepackage{bm}
\usepackage{physics}
\usepackage{tikz}
\usepackage{multirow}
\usepackage{xcolor}

\begin{document}

\preprint{APS/123-QED}

\title{Intrinsic spin Hall effect in topological insulators: A first-principles study}

\author{S. M. Farzaneh}
\email{farzaneh@nyu.edu}
 \affiliation{Department of Electrical and Computer Engineering, New York University, Brooklyn, New York 11201, USA}

\author{Shaloo Rakheja}
\email{rakheja@illinois.edu}
 \affiliation{Holonyak Micro and Nanotechnology Laboratory University of Illinois at Urbana-Champaign, Urbana, Illinois 61801, USA}

\begin{abstract}
The intrinsic spin Hall conductivity of typical topological insulators Sb$_2$Se$_3$, Sb$_2$Te$_3$, Bi$_2$Se$_3$, and Bi$_2$Te$_3$ in the bulk form, is calculated from first-principles by using density functional theory and the linear response theory in a maximally localized Wannier basis.  
The results show that there is a finite spin Hall conductivity of 100--200 ($\hbar$/2e)(S/cm) in the vicinity of the Fermi energy. 
Although the resulting values are an order of magnitude smaller than that of heavy metals, they show a comparable spin Hall angle due to their relatively lower longitudinal conductivity.
The spin Hall angle for different compounds are then compared to that of recent experiments on topological-insulator/ferromagnet heterostructures. 
The comparison suggests that the role of the bulk in generating a spin current and consequently a spin torque in magnetization switching applications is comparable to that of the surface including the spin-momentum locked surface states and the Rashba-Edelstein effect at the interface. 
\end{abstract}

\maketitle

\section{Introduction}
A topological state of matter is distinguished by its insulating bulk, but conducting surface states that are robust against disorder. 
The surface states in a topological insulator carry opposite spins while propagating in opposite directions~\cite{hsieh_tunable_2009}. 
Due to their strong spin-orbit coupling,
topological insulators are capable of switching an adjacent thin-film ferromagnet in a bilayer structure without the need to apply any external magnetic fields~\cite{mellnik_spin-transfer_2014}.
Both the bulk and the surface states are reported to be involved in generating an electric-field-induced spin torque on the ferromagnet \cite{kondou_fermi-level-dependent_2016}.
The spin polarization on the surface is due to the spin-momentum locking \cite{hsieh_tunable_2009, wang_room_2017} of the surface states as well as the Rashba-Edelstein effect \cite{edelstein_spin_1990, zhu_rashba_2011} at the interface with the ferromagnet, while the bulk of the topological insulator contributes via the intrinsic spin Hall effect \cite{seifert_spin_2018, liu_direct_2018}. 
Several experimental studies \cite{kondou_fermi-level-dependent_2016, wang_room_2017, wu_room-temperature_2019} have attempted to distinguish the contribution of the surface and the bulk states through magnetization switching in topological-insulator/ferromagnet heterostructures. 
However, theoretical estimates of the intrinsic bulk contribution are limited. 
In this work, we quantify the role of the bulk in spin generation by calculating the intrinsic spin Hall conductivity of four topological insulators namely Sb$_2$Se$_3$, Sb$_2$Te$_3$, Bi$_2$Se$_3$, and Bi$_2$Te$_3$, by using first-principles calculations. 
These materials, along with their alloys, are of the first experimentally realized \cite{hsieh_tunable_2009, chen_experimental_2009, roushan_topological_2009, xia_observation_2009} three dimensional topological insulators and have been studied more extensively especially in terms of the spin Hall effect.

The spin Hall effect is the accumulation of spin on the surface of a material in response to an applied electric field.
Dyakonov and Perel \cite{dyakonov_possibility_1971} introduced the idea of generating spin polarization with a charge current via the Mott scattering which is a spin-dependent scattering off a Coulomb potential in the presence of spin-orbit coupling. 
They introduced a phenomenological \textit{spin electric coefficient} term, which models the generation of a transverse spin current via an external electric field. 
The scattering potential by impurities and phonons can also result in an asymmetric scattering cross section leading to the accumulation of spin, an effect called the \textit{extrinsic} spin Hall effect \cite{hirsch_spin_1999, kato_observation_2004}.
However, it has been shown \cite{wunderlich_experimental_2005} that even in the absence of extrinsic effects, spin accumulation occurs due to the finite spin-orbit coupling of the underlying crystal.
Therefore, the contribution of non-zero orbital angular momentum in the Bloch wavefunctions along with an external electric field gives rise to a spin accumulation, a phenomenon known as the \textit{intrinsic} spin Hall effect \cite{murakami_spin-hall_2004, sinova_universal_2004}. 

Topological insulators are a distinct state of quantum matter where the band structure is topologically different than that of ordinary/trivial insulators due to inverted bands at the Fermi level. 
Although insulating in the bulk, they are conducting on their surfaces where they meet a trivial insulator that is the vacuum \cite{xia_observation_2009, zhang_topological_2009}. Recently, topological insulators have been used to generate spin current to electronically switch the magnetization of a proximal thin-film magnet~\cite{mellnik_spin-transfer_2014}.
Compared to heavy metals, such as platinum~\cite{kimura_room-temperature_2007} and tantalum~\cite{liu_spin-torque_2012}, which are also used as spin-current generators, topological insulators are expected to consume less energy while yielding the same spin Hall angle \cite{liu_direct_2018}, which is beneficial in realizing low power spintronics.

In this work we focus on the intrinsic ability of the bulk of topological insulators in generating spin currents through the spin Hall effect. 
We calculate the spin Hall conductivity of the four compounds (Sb/Bi)$_2$(Se/Te)$_3$ from first principles, that is by solving the Schr\"{o}dinger equation in the framework of density functional theory and using the solution to calculate the linear response coefficients. 
Previous theoretical studies on the intrinsic spin Hall conductivity in topological insulators are limited to those of HgTe \cite{matthes_intrinsic_2016} by using first principles and Bi$_{1-x}$Sb$_{x}$ \cite{sahin_tunable_2015} and Bi$_2$Se$_3$ \cite{peng_spin_2016, liu_spin-polarized_2015} by using a tight binding and an effective Hamiltonian.
It is worth noting that the calculation of spin Hall conductivity requires integration over the entire Brillouin zone and that all the bands below the Fermi energy contribute toward the spin Hall effect. Hence, the effective Hamiltonian, which is valid only in the vicinity of the Fermi energy, may not provide an adequate description of the spin Hall conductivity. First-principles calculations are therefore necessary to accurately quantify the intrinsic strength of spin generation in typical topological insulators, as well as understanding the bulk contribution in magnetization switching applications. 
We furthermore provide estimates of the spin Hall angle in different compounds using the large body of experimental studies on the topological-insulator/ferromagnet heterostructures. 

Section \ref{sec:exp} provides an overview of the experimental studies and techniques in estimating the spin Hall angle. 
Theoretical details of calculating the spin Hall conductivity are given in Section \ref{sec:she}. 
The symmetry properties of the crystal and the spin Hall conductivity tensor are discussed in Section~\ref{sec:symm}.
Results are presented in Section~\ref{sec:result} along with a comparison of different values of the spin Hall angle reported in the literature. 
Section \ref{sec:conclusion} concludes the paper. 
Computational details and the first-principles setup are presented in the Appendix.

\section{\label{sec:exp}Review of Experiments}
The ability of a material to generate a spin current via the spin Hall effect is measured by the spin Hall angle (efficiency) $\theta$ which is proportional to the ratio of the spin current density $J_\alpha^\gamma$ to the charge current density $J_\beta$, i.e., $\theta = (2e/\hbar)J_\alpha^\gamma/J_\beta = (2e/\hbar)\sigma_{\alpha\beta}^\gamma/\sigma_{\beta\beta}$ where $\sigma_{\alpha\beta}^\gamma$ is the spin Hall conductivity and $\sigma_{\beta\beta}$ is the longitudinal charge conductivity. 
The spin Hall angle is usually measured via three different techniques namely the spin-torque ferromagnetic resonance (ST-FMR), the second harmonic Hall voltage, and the helicity-dependent photoconductance. 
These methods are briefly introduced in this section. In section \ref{sec:result}, we show the spread of experimentally reported values of $\theta$ in various topological materials using different measurement schemes.

The ST-FMR technique was introduced by Liu et al. \cite{liu_spin-torque_2011} in the context of spin Hall effect in heavy-metal/ferromagnet heterostructures such as platinum/permalloy bilayers.
Later, Mellnik et al. \cite{mellnik_spin-transfer_2014} utilized this technique for topological-insulator/ferromagnet heterostructures. 
The ST-FMR
method is based on the spin-torque driven magnetization resonance when a radio frequency (RF) charge current flows in the proximal charge-to-spin convertor, i.e. heavy metal or topological insulator. Additionally, a large constant magnetic field causing the magnetic order to precess, is also applied.
Based on the solution to the Landau-Lifshitz-Gilbert equation describing the magnetization dynamics, the magnetoresistance of the structure is expressed as a linear combination of a symmetric and an anti-symmetric Lorentzian function with respect to the external magnetic field. 
The symmetric Lorentzian describing the contribution of the spin current density to the spin torque is used to quantify the spin Hall angle. 
The ST-FMR technique has been used in several experiments to demonstrate magnetization switching in topological-insulator/ferromagnet heterostructures~\cite{mellnik_spin-transfer_2014, jamali_giant_2015, wang_topological_2015, kondou_fermi-level-dependent_2016, han_room-temperature_2017, wang_room_2017, fanchiang_strongly_2018}.

The second harmonic technique was introduced by Garello et al. \cite{garello_symmetry_2013} to measure spin-orbit torques in ferromagnetic materials and was modified by Fan et al. \cite{fan_magnetization_2014} to measure spin-transfer torque in topological-insulator/ferromagnet heterostructures.
The setup of the second harmonic technique is similar to that of the ST-FMR in that the spin torque acting on the ferromagnet results from the charge-to-spin conversion in a proximal heavy metal or topological insulator layer. However, the magnetic field that is used in the second harmonic method is not static, but rotates in a plane perpendicular to the sample and parallel to the RF charge current. 
The first frequency component of the resulting Hall voltage in the sample is proportional to the RF current with the Hall resistance as the proportionality constant. The second harmonic component is shown to be proportional to the spin-transfer torque with a proportionality constant that depends on the anomalous Hall coefficient and the relative orientation of the magnetic field and the magnetization. 
Experimental works that utilize this measurement technique to quantify spin Hall angle include~\cite{fan_magnetization_2014, yasuda_current-nonlinear_2017, wu_room-temperature_2019}.

The photoconductive method has only recently been utilized to study spin Hall effect in topological insulators \cite{liu_direct_2018, seifert_spin_2018}. 
Unlike the ST-FMR and the second harmonic method, the photoconductive method does not rely on the presence of a coupled ferromagnetic thin film to measure the spin Hall angle. In this method, spin accumulation that appears on the lateral edges of the sample due to a charge current is directly probed. That is, by shining a laser light with modulated helicity, the population of the spins can be locally changed.
This change in the population of the spins reflects in a transverse spin-dependent voltage, which is shown to be proportional to the spin Hall angle with a proportionality constant that depends on the material geometry and transport properties such as sample resistivity. 
Therefore, by measuring the helicity-dependent photovoltage across the sample one can extract the value of the spin Hall angle.

\section{\label{sec:she}Spin Hall Conductivity}
In the linear response theory, the spin conductivity $\sigma_{\alpha\beta}^\gamma$ is a tensor that connects the applied electric field $E_\beta$ to the a spin current density $J_\alpha^\gamma$ which is the response of the system, i.e., $J_\alpha^\gamma = \sigma_{\alpha\beta}^\gamma E_\beta$. 
The spin \textit{Hall} conductivity is then a component in which $\alpha$ and $\beta$ are perpendicular to each other. 
Utilizing the Kubo formula, the spin conductivity can be written in terms of a Berry-like curvature $\Omega_{\alpha\beta,n}^\gamma(\vb*{k})$, also called the spin Berry curvature, as follows \cite{gradhand_first-principle_2012}
\begin{equation}
\label{eq:shc}
    \sigma_{\alpha\beta}^\gamma = -\qty(\frac{e^2}{\hbar})\qty(\frac{\hbar}{2e})\int\frac{d^3\vb*{k}}{(2\pi)^3}\sum_{n} f(\epsilon_{n,\vb*{k}})\Omega_{\alpha\beta,n}^\gamma(\vb*{k}),
\end{equation}
where $f(\epsilon_{n,\vb*{k}})$ is the Fermi-Dirac distribution function. 
The spin Berry curvature is given as 
\begin{equation}
\label{eq:berry}
    \Omega_{\alpha\beta,n}^\gamma(\vb*{k}) = \hbar^2\sum_{m\not=n} \frac{-2\Im{\bra{n\vb*{k}}\mathcal{J}_{\alpha}^\gamma\dyad{m\vb*{k}} v_{\beta}\ket{n\vb*{k}}}}{(\epsilon_{n,\vb*{k}} - \epsilon_{m,\vb*{k}})^2}, 
\end{equation}
where $v_\beta$ is the velocity operator and  $\mathcal{J}_{\alpha}^\gamma = \acomm{v_\alpha}{\sigma_\gamma}/2=(v_\alpha\sigma_\gamma + \sigma_\gamma v_\alpha)/2$ is the spin velocity operator. 
The structure of the spin Berry curvature suggests that the spin Hall effect can be viewed as an intermixture of the Bloch functions in the $k$ space. It is worth noting that the integral of this curvature over the $k$ space is not quantized \cite{matthes_intrinsic_2016, murakami_spin-hall_2004}, unlike the Berry curvature in integer quantum Hall systems. 
Based on first-principles calculations, Eq. (\ref{eq:shc}) has been used to study the spin Hall effect in semiconductors \cite{guo_ab_2005, yao_sign_2005, feng_intrinsic_2012} and heavy metals \cite{guo_intrinsic_2008, qiao_calculation_2018, zhou_intrinsic_2019}. 
Previous calculations on topological insulators have been performed for HgTe \cite{matthes_intrinsic_2016} based on first principles and also for Bi$_{1-x}$Sb$_x$ \cite{sahin_tunable_2015} and Bi$_{2}$Se$_3$ \cite{peng_spin_2016, liu_spin-polarized_2015} based on a tight binding and an effective Hamiltonian, respectively. 
However, first-principles calculations of the spin Hall conductivity for (Bi/Sb)$_2$(Se/Te)$_3$ crystals have not been reported previously.

We evaluate Eq. (\ref{eq:shc}) based on first principles for (Bi/Sb)$_2$(Se/Te)$_3$ crystals. Although this equation can be evaluated directly from the Bloch functions, a more computationally efficient method involves the Wannier functions instead.
This method was introduced by Wang et al. \cite{wang_abinitio_2006} for the calculation of anomalous Hall conductivity. 
Later, the Wannier method was used to calculate the spin Hall conductivity of transition metal dichalcogenides \cite{feng_intrinsic_2012} and $\alpha$-Ta and $\beta$-Ta \cite{qiao_calculation_2018}. 
The Wannier method is based on maximally localized Wannier functions \cite{marzari_maximally_1997} which are constructed by a unitary gauge transformation of the Bloch basis.
The corresponding unitary matrices are obtained via an optimization scheme in which the spread of the real space Wannier functions is minimized iteratively.
Due to the gauge freedom in the Bloch basis, the Bloch functions calculated from first principles are generally highly discontinuous in the $k$ space. 
Since the Wannier functions are related to the Bloch functions through the Fourier transform, the gauge freedom is also present in the Wannier basis. 
However, by maximally localizing the Wannier functions, it is possible to find a gauge that provides the most smooth Bloch basis which enables the evaluation of the Brillouin zone integral over a fine mesh through an efficient interpolation scheme. 
Recently, Qiao et al. \cite{qiao_calculation_2018} have implemented this scheme in a module of the Wannier90 \cite{pizzi_wannier90_2020} open source code which we utilize to calculate the spin Hall conductivity of topological insulators. 

We briefly review the procedure for evaluating Eq. (\ref{eq:shc}). 
First, the band structure is calculated on a relatively coarse mesh in the plane wave basis. The resulting Bloch functions are then projected onto Wannier functions. The Wannier functions are maximally localized via an optimization process which provides the optimal unitary transformation between the Bloch basis and the Wannier basis. The matrix elements are then calculated in the Wannier basis. And finally, the spin Berry curvature is integrated on the whole Brillouin zone on a fine mesh via interpolation. 

Although the spin conductivity is a 27-component tensor, not all components need to be evaluated separately. 
Recently, symmetry was utilized to determine the non-zero components of the spin conductivity tensor and simplify the calculations in MoS$_2$ and WTe$_2$ \cite{zhou_intrinsic_2019} and TaAs family of Weyl semmimetals \cite{sun_strong_2016}. 
In the next section, based on the work of Seemann et al. \cite{seemann_symmetry-imposed_2015}, we discuss the symmetry of (Bi/Sb)$_2$(Se/Te)$_3$ crystals and show how it simplifies the spin conductivity tensor where many components become zero and the rest are not all independent.

\section{\label{sec:symm}Crystal Structure and Symmetries}
\begin{figure}[ht]
  \centering
  \includegraphics[width=1.0\linewidth]{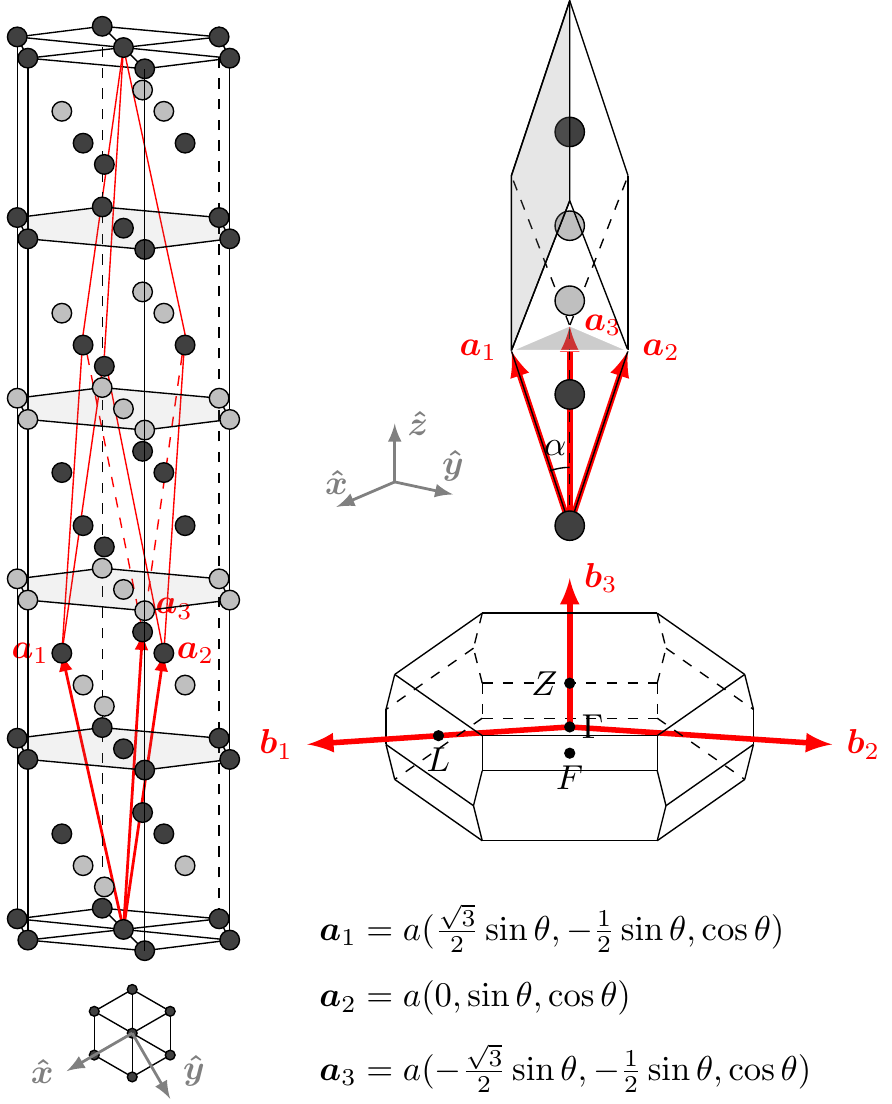}
  \caption{The conventional and the primitive unit cell of the trigonal crystal of the four binary compounds (Bi/Sb)$_2$(Se/Te)$_3$ where Bi or Sb elements are denoted by light circles and Se or Te elements by dark circles. 
  The primitive vectors are drawn in red where $\alpha$ is the angle between each pair and $\theta$ is the polar angle. 
  The first Brillouin zone along with the reciprocal vectors and the special points are shown on the right.}
  \label{fig:crystal}
\end{figure}

The four binary compounds (Bi/Sb)$_2$(Se/Te)$_3$ share the same crystal structure which is classified as the trigonal crystal with a single three-fold high symmetry axis. 
The symmetry of the crystal is described by the symmorphic space group $R\overline{3}m$ (\#166). 
Figure \ref{fig:crystal} illustrates the conventional hexagonal unit cell and the primitive rhombohedral unit cell of these compounds along with the first Brillouin zone. 
There are five atoms per unit cell denoted in the figure. The symmetry of the crystal can predict several properties of the system such as the number of degeneracies and the selection rules to identify the components of the linear response tensor that evaluate to zero.

For instance, the group of the wavevector at the $\Gamma$ point of the Brillouin zone is homomorphic to the point group $D_{3d}$, and, therefore, the energy levels at the $\Gamma$ point are labeled by the irreducible representation of the double group for group $D_{3d}$. 
Since these irreducible representations are two-dimensional~\cite{ dresselhaus_group_2008, aroyo_crystallography_2011}, there are only two-fold degeneracies at the $\Gamma$ point which represent the Kramers doublets and are protected by the time-reversal symmetry. 
The group of the wavevector at other points of the Brillouin zone is a subgroup of $D_{3d}$, and, therefore, 
the bands do not merge into degenerate ones anywhere in the Brillouin zone.
It is possible that accidental degeneracies appear at a number of points in the Brillouin zone where some energy levels get close to each other. These points lead to spikes in the Kubo formula, where the energy difference between different levels appears in the denominator and could make the integration over the Brillouin zone challenging. 

Symmetry restricts the spin conductivity tensor through selection rules which determine the zero matrix elements of a given operator. 
These symmetry properties of the spin conductivity tensor have been worked out by Seemann et al. \cite{seemann_symmetry-imposed_2015} based on an earlier method by Kleiner \cite{kleiner_space-time_1966}.
Within this method, which takes the Kubo formula as the starting point, the components of the spin conductivity tensor are derived in terms of each other through the symmetry elements of the space group of the underlying crystal. 
Depending on the magnetic space group classification, i.e. the magnetic Laue group, the general form of the tensor changes. 

The crystal structure of (Bi/Sb)$_2$(Se/Te)$_3$ which is described by the space group $R\overline{3}m$ corresponds to the non-magnetic Laue group $\overline{3}m11'$.
The spin conductivity tensor is denoted by $\sigma_{\alpha\beta}^\gamma$ where $\alpha,\beta$, and $\gamma$ represent the direction of the spin current, the direction of the electric field, and the spin polarization, respectively. 
The general form of this tensor according to the symmetry restrictions is as follows~\cite{seemann_symmetry-imposed_2015} 
\begin{equation}
    \vb*{\sigma}^x = 
    \begin{pmatrix}
    0 & \sigma_{xx}^y & 0 \\
    \sigma_{xx}^y & 0 & -\sigma_{xz}^y\\
    0 & -\sigma_{zx}^y & 0 
    \end{pmatrix},
\end{equation}
\begin{equation}
    \vb*{\sigma}^y = 
    \begin{pmatrix}
    \sigma_{xx}^y & 0 & \sigma_{xz}^y \\
    0 & -\sigma_{xx}^y & 0\\
    \sigma_{zx}^y & 0 & 0 
    \end{pmatrix},
\end{equation}
\begin{equation}
    \vb*{\sigma}^z = 
    \begin{pmatrix}
    0 & \sigma_{xy}^z & 0 \\
    -\sigma_{xy}^z & 0 & 0\\
    0 & 0 & 0 
    \end{pmatrix}\cdot
\end{equation}

As seen from the above equations, several components are zero, while not all non-zero components are independent.
For instance, $\sigma_{zy}^x = -\sigma_{zx}^y$. 
There are only four independent components of the tensor namely $\sigma_{xx}^y$, $\sigma_{xz}^y$, $\sigma_{zy}^x$, and $\sigma_{xy}^z$. 
Here, we calculate all these four non-zero components of the spin conductivity tensor.
The majority of the magnetization switching experiments involve the spin current in the $[111]$ direction ($z$ direction here, i.e., $\sigma_{zy}^x$ component) which is referred to spin Hall conductivity herein unless otherwise stated.
We note that since we are dealing with non-magnetic space groups which contain the time-reversal operator as a group element, the Onsager reciprocity relations are satisfied~\cite{seemann_symmetry-imposed_2015}.
As a consequence, the \textit{inverse} spin Hall conductivity has an equal magnitude to that of the direct one. 

\section{\label{sec:result}Results \& Discussions}
\begin{figure}
    \includegraphics[width=1.0\linewidth]{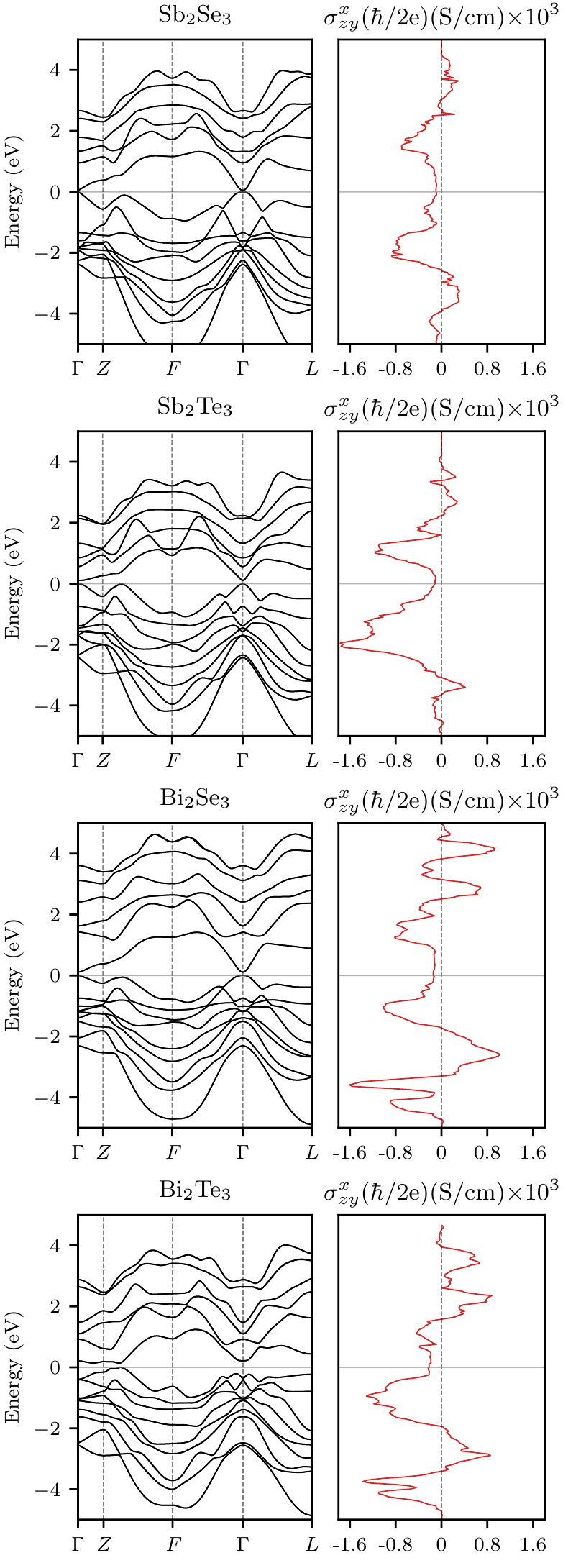} 
    \caption[example] 
    {The band structure (left) and the spin Hall conductivity (right) of the four topological insulators Sb$_2$Se$_3$, Sb$_2$Te$_3$, Bi$_2$Se$_3$, and Bi$_2$Te$_3$. 
    The energy axis is relative to the Fermi energy denoted by the gray horizontal line.}
    \label{fig:band} 
\end{figure}  
First-principles calculations within the density functional theory are performed to obtain the band structure and the spin Hall conductivity of the four topological insulators Sb$_2$Se$_3$, Sb$_2$Te$_3$, Bi$_2$Se$_3$, and Bi$_2$Te$_3$. 
The details of the first-principles setup and the calculation of spin Hall conductivity are presented in the Appendix. 
Figure \ref{fig:band} illustrates the band structure of these four compounds on the left column as well as their corresponding spin Hall conductivity, $\sigma_{zy}^x$, as a function of the energy level on the right column. 
Both the band structure and the spin Hall conductivity are obtained in the Wannier basis.
The bands shown in the energy window of the figure are composed of the $s$ and $p$ orbitals. 
The fully occupied $d$ orbitals make highly narrow bands far below the Fermi level and, therefore, their contributions to the spin Hall effect is negligible. 
However, they are included in the pseudopotentials used for the first-principles calculations of the Kohn-Sham orbitals. 

As seen from Fig.~\ref{fig:band}, the spin Hall conductivity is non-zero, albeit small compared to that of heavy metals, and is constant inside the gap and also to some extent beyond the gap. This suggests that the bulk in topological insulators could generate a finite spin current even when the Fermi level is located inside the gap and at zero temperature. 
Due to their finite spin Hall conductivity and limited longitudinal charge conductivity, the spin Hall angle of bulk topological insulators is comparable to that of heavy metals. Therefore, bulk topological insulators are an excellent candidate for energy-efficient charge-to-spin conversion in spin-based devices.

The spin Hall conductivity in all of the four compounds studied here show mostly a similar dependence on the Fermi energy. 
That is, the spin Hall conductivity has a constant magnitude inside the energy gap and has peaks at certain energy levels where bands get too close to each other and result in accidental degeneracies. 
The values of spin Hall conductivity of Sb$_2$Se$_3$, Sb$_2$Te$_3$, Bi$_2$Se$_3$, and Bi$_2$Te$_3$ at the Fermi level are 93.8, 113, 147, and 218 ($\hbar$/2e)(S/cm), respectively. 
These values are of the same order of magnitude as the ones reported in the literature for slightly different materials by different methods \cite{matthes_intrinsic_2016, sahin_tunable_2015}. 
As one goes from low spin-orbit strength of Sb$_2$Se$_3$ to the relatively higher spin-orbit strength of Bi$_2$Te$_3$, the magnitude of the spin Hall conductivity at the Fermi level increases monotonically. 

\begin{figure}
    \includegraphics[width=1.0\linewidth]{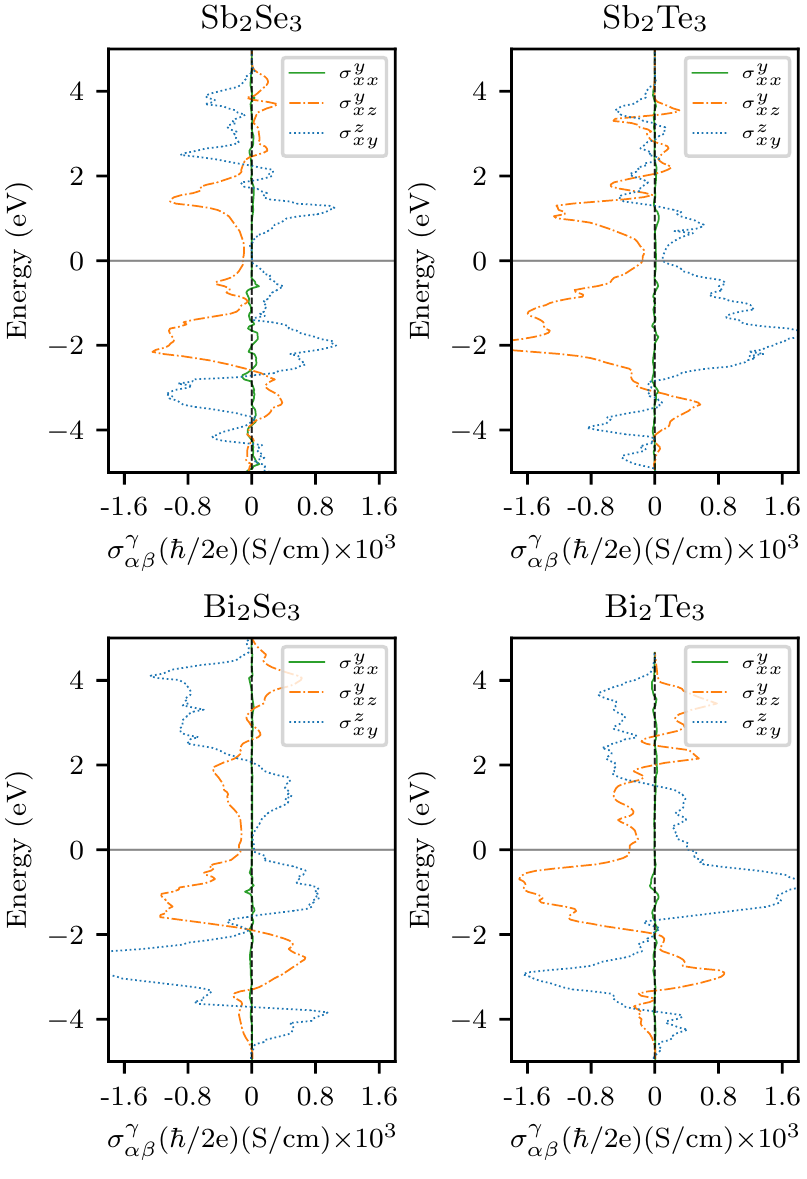} 
    \caption[] 
    {Other components of the spin conductivity tensor allowed by the symmetry namely $\sigma_{xx}^y$ and $\sigma_{zy}^x$. The $\sigma_{xx}^y$ component is negligible.}
    \label{fig:shc-other} 
\end{figure}

There are three other non-zero components of the spin conductivity tensor, allowed by the symmetry, namely $\sigma_{xx}^y$, $\sigma_{xz}^y$, and $\sigma_{xy}^z$. 
Figure \ref{fig:shc-other} illustrates the values of these components over the energy.
These components are associated with in-plane spin currents which cause the spin to accumulate on the lateral edges of the sample.
As seen from the figure, the values of the $\sigma_{xx}^y$ component are minute for all the four compounds.  
The values of $\sigma_{xz}^y$ and $\sigma_{xy}^z$ components at the Fermi energy are, respectively, 113 and 0 for Sb$_2$Se$_3$, 154 and 100 for Sb$_2$Te$_3$, 162 and 27.2 for Bi$_2$Se$_3$, and 314 and 406 for Bi$_2$Te$_3$, all in units of ($\hbar$/2e)(S/cm). 
Although the majority of the experimental works involve the magnetization switching via the out-of-plane spin current, i.e., $\sigma_{zy}^x$, the three other in-plane components also show comparable values. 
This has a consequence in an experimental setup. 
For instance, in a typical setup where there is usually a substantial perpendicular electric field ($z$ direction), either from the substrate or the gates, the component $\sigma_{xz}^y$ leads to a spin accumulation at the lateral edges of the sample.

\begin{figure}
    \includegraphics[width=1.0\linewidth]{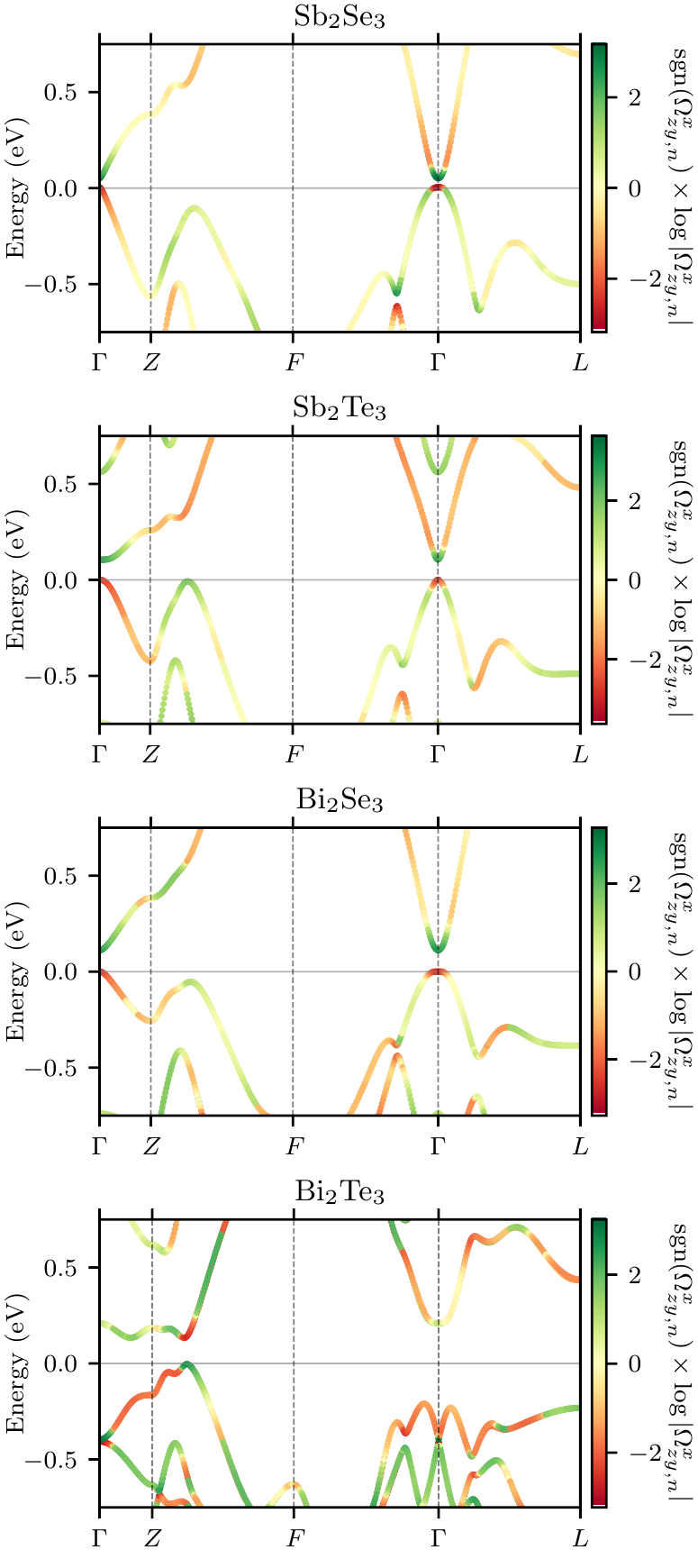} 
    \caption[] 
    {The band-projected spin-Berry curvature, $\Omega_{zy,n}^x(\vb*{k})$, in the vicinity of the Fermi energy for Sb$_2$Se$_3$, Sb$_2$Te$_3$, Bi$_2$Se$_3$, and Bi$_2$Te$_3$. 
    The energy axis is relative to the Fermi energy denoted by the gray horizontal line.}
    \label{fig:berry} 
\end{figure}

A more detailed insight into the origin of the finite spin Hall conductivity inside the gap can be gained by studying the contributions of the bands in the vicinity of the Fermi energy. 
Figure \ref{fig:berry} depicts the band projected spin-Berry curvature, i.e., $\Omega_{zy,n}^x(\vb*{k})$, for energies close to the band gap.
The bands are colored by the sign and magnitude of the spin-Berry curvature, i.e., $\text{sgn}(\Omega_{zy,n}^x(\vb*{k}))\log|\Omega_{zy,n}^x(\vb*{k})|$
The conduction and the valence bands show large contributions to the spin-Berry curvature, especially close to the Fermi energy. 
Moreover, the sign of $\Omega_{zy,n}^x(\vb*{k})$ flips suddenly as the energy nears the band gap. 
The strong magnitude of the spin-Berry curvature and its sudden sign flip at the Fermi energy suggests that the spin Hall conductivity is related to the topological order of the bands.  
We also note that previous calculations on trivially gapped semiconductors \cite{yao_sign_2005} show a residual finite spin Hall conductivity inside the gap, but its origin does not seem to be topological.

\bgroup
\def\arraystretch{1.5}
\begin{table*}[ht]
\centering
\caption{Comparison of spin Hall conductivity $\sigma_{zy}^x$ and the spin Hall angle (efficiency), $\theta =(2e/\hbar)\sigma_{zy}^x/\sigma_{yy}$, from first-principles calculations and experimental observations.}
\label{tb:spin-hall-angle}
\begin{tabular}{ccccccr|cc}
\hline
& & & & \multicolumn{3}{c}{Experiments} & \multicolumn{2}{c}{ab initio (this work)}\\
Ref. & Material & Method & thickness nm & $|\sigma_{zy}^x|$ ($\hbar$/2e)(S/cm) &  $\theta$ & $\sigma_{yy}$ (S/cm) & $|\sigma_{zy}^x|$ & $\theta$ \\
\hline 
\onlinecite{mellnik_spin-transfer_2014} & Bi$_2$Se$_3$ & ST-FMR & $8$ & $1.1 - 2.0 \times 10^3$ & $2.0 - 3.5$ & $5.7\times 10^2$ & $1.47\times 10^2$ & $0.26$\\
\onlinecite{fan_magnetization_2014} & (Bi$_{0.5}$Sb$_{0.5}$)$_2$Te$_3$ & 2nd Harmonic & $3$ & & $180-425$ & $2.227\times 10^2$ & *$1.77\times 10^2$ & $0.79$\\
\onlinecite{jamali_giant_2015} & Bi$_2$Se$_3$ & ST-FMR & $5-10$ & & $0.02-0.34$ & $1\times 10^3$ & $1.47\times 10^2$ & $0.15$\\
\onlinecite{wang_topological_2015} & Bi$_2$Se$_3$ & ST-FMR & 20 & & $0.1$ & $2.50\times 10^3$ & $1.47\times 10^2$ & $0.06$ \\
\onlinecite{kondou_fermi-level-dependent_2016} & Bi$_2$Te$_3$ & ST-FMR & 8 & $4.0\times 10^3$ & $1.0$ & $3.75\times 10^3$ & $2.18\times 10^2$ & $0.06$\\
\onlinecite{han_room-temperature_2017} & (Bi,Sb)$_2$Te$_3$ & Hall resistance & $8.0$ & & $0.4$ & $2.488\times 10^2$ & *$1.77\times 10^2$ & $0.71$ \\
\onlinecite{han_room-temperature_2017} & Bi$_2$Se$_3$ & Hall resistance & $7.4$ & & $0.16$ & $9.434\times 10^2$ & $1.47\times 10^2$ & $0.15$ \\
\onlinecite{wang_room_2017} & Bi$_2$Se$_3$ & ST-FMR & $20$ & & $0.3$ & $1.0\times 10^3$ & $1.47\times 10^2$ & $0.15$ \\
\onlinecite{yasuda_current-nonlinear_2017} & (Bi$_{1-x}$Sb$_{x}$)$_2$Se$_3$ & 2nd Harmonic & $5$ & & $160$ & $2.44\times 10^2$ & *$1.09\times 10^2$ & $0.45$\\
\onlinecite{liu_direct_2018} & Bi$_2$Se$_3$ & Photoconductance & $9$ & & $0.0085$ & $1.18\times 10^3$ & $^\dagger 2.72\times 10^1$ & $0.023$ \\
\onlinecite{wu_room-temperature_2019} & Bi$_2$Te$_3$ & 2nd Harmonic & 6 & & $0.08$ & $1.5\times 10^3$ & $2.18\times 10^2$ & $0.15$ \\
 & Sb$_2$Se$_3$ & & & & & & $9.38\times 10^1$ & \\
 & Sb$_2$Te$_3$ & & & & & & $1.13\times 10^2$ & \\
\hline
\end{tabular}\\
\footnotesize{$*$ These values of $\sigma_{zy}^x$ for alloys are obtained by a weighted average over that of non-alloy compounds.\\
$\dagger$ This value reflects the $\sigma_{xy}^z$ component of the spin conductivity tensor.} 
\end{table*}
\egroup

We summarise the recent experimental works on spin Hall effect in topological insulators in Table \ref{tb:spin-hall-angle}. For each experiment mentioned in this table, several properties are listed such as the material, the measurement method, the thickness of the topological insulator, the magnitude of the spin Hall conductivity (if reported), $|\sigma_{zy}^x|$, the spin Hall angle, $\theta$, and the longitudinal charge conductivity $\sigma_{yy}$. 
The first-principles results in this work are listed in the last two columns where the magnitude of the spin Hall conductivity $|\sigma_{zy}^x|$ is obtained at the Fermi energy and the value of the spin Hall angle is estimated by using the longitudinal conductivity $\sigma_{yy}$ corresponding to each experiment, i.e., $\theta = (2e/\hbar)|\sigma_{zy}^x|/\sigma_{yy}$. 
For the non-stoichiometric compounds we report only estimates of the first-principles spin Hall conductivity by taking a weighted average over that of the stoichiometric ones. 
We note that since there are no experimental data available for Sb$_2$Se$_3$ and Sb$_2$Te$_3$ compounds, only the first-principles results are reported. 
It should be noted that, the photoconductive experiment measures a different component of the spin conductivity that is $\sigma_{xy}^z$. Therefore, the corresponding first-principles values of $\sigma_{xy}^z$ are reported instead. 

\begin{figure}
    \includegraphics[width=1.0\linewidth]{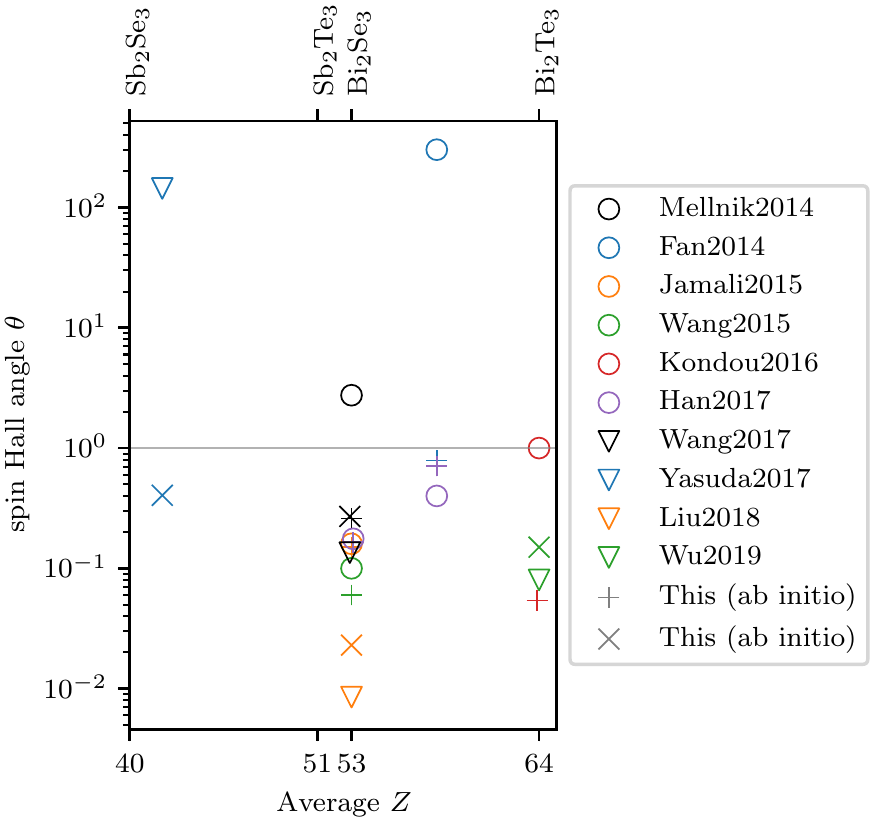} 
    \caption[] 
    {Comparison of spin Hall angle of various bulk topological insulators reported experimentally and computed in this work from first principles. The horizontal axis specifies the average atomic number of the atoms in the unit cell.}
    \label{fig:spin-hall-angle} 
\end{figure}  

To compare different experimental techniques and to put the reported values of spin Hall angle into perspective, Table \ref{tb:spin-hall-angle}'s data are plotted in Fig.~\ref{fig:spin-hall-angle}.
This figure shows the spin Hall angle of different crystals versus the average atomic number of their unit cell. 
In this figure the experimental data points are denoted by circles and triangles  whereas, our estimates of the spin Hall angle based on first-principles calculations are denoted by $+$ and $\times$ symbols with the same color as their corresponding experimental value reported in other works. 
As seen from the figure the majority of the data points lie in the $0.1<\theta<1.0$ range which is comparable to the values reported for the heavy metals such as $0.056<\theta<0.16$ for platinum \cite{liu_spin-torque_2011} and $0.12<\theta<0.15$ for tantalum \cite{liu_spin-torque_2012}. 
As mentioned previously, this suggests that for the same spin current, topological insulators require a lower value of charge current compared to heavy metals, which possess a relatively higher conductivity by more than an order of magnitude. 

The first-principles results show a reasonable match with most data points especially for Bi$_2$Se$_3$ which has several data points. 
However, two of the data points \cite{fan_magnetization_2014, yasuda_current-nonlinear_2017} related to the second harmonic Hall voltage method are orders of magnitude higher than that of other methods. 
This large discrepancy is related to the measurement of the spin-orbit torques. 
Generally, there are two different spin torques, field-like and damping-like torques, that affect the 2nd harmonic voltage. 
The spin Hall angle is related to the damping-like torque (which is caused by the spin current). 
In studies \cite{fan_magnetization_2014, yasuda_current-nonlinear_2017} that report a very large spin-torque ratio, the contribution of the field-like torque in the Hall voltage is not taken into account. 
However, in a recent study \cite{wu_room-temperature_2019}, which makes a distinction between the two torques and measures the damping-like torque separately, it was shown that the resulting spin-torque ratio is comparable to the spin Hall angle reported by other techniques such as ST-FMR method. 
Another possible reason for this discrepancy could be magnon scattering which incidentally was reported to be negligible in Ref. \cite{wu_room-temperature_2019}.
As mentioned by Yasuda et al. \cite{yasuda_current-nonlinear_2017}, in certain configurations, the second harmonic methods tend to overestimate the spin Hall angle because the nonlinearity of the transverse Hall voltage is dominated by an asymmetric magnon scattering and not by the spin-orbit torque. 
The photoconductive experiments \cite{liu_direct_2018} provide a lower bound because no interface or ferromagnetic effects are present and therefore, the reported spin Hall angle could be attributed to the intrinsic spin Hall effect in the bulk. 
Since, the first principles results also reflect only the bulk contribution, one expects a better match with the photoconductive data points than with the other methods. 
However, in the case of Bi$_2$Se$_3$, the first-principles estimate of the spin Hall angle $0.023$ is higher, by a factor of 3, than the value reported by Liu et al. \cite{liu_direct_2018} that is $0.0085$. 
This discrepancy could be related to the premises in Ref. \cite{liu_direct_2018} where it is assumed that the spin accumulation at the lateral edges of the sample, due to a longitudinal charge current in the $y$-direction, are $z$-polarized because of the $\sigma_{xy}^z$ component. 
However, based on the symmetry study in Sec. \ref{sec:symm}, in the presence of an electric field in the $z$ direction, there is another component $\sigma_{xz}^y$ with a larger contribution to the spin accumulation at the edges of the sample which could have affected the measured photovoltage in Ref. \cite{liu_direct_2018}.

\section{\label{sec:conclusion}Conclusions}
First-principles calculations of the spin Hall conductivity of typical topological insulators Sb$_2$Se$_3$, Sb$_2$Te$_3$, Bi$_2$Se$_3$, and Bi$_2$Te$_3$ show finite values at the Fermi energy. 
These values are lower by an order of magnitude than those of heavy metals. However, due to their relatively low current conduction capability, the spin Hall angle in topological insulators is comparable to that reported in heavy metals. 
We compare theoretical results against experimental observations of spin Hall angle via direct helicity-dependent photovoltage measurements as well as in bilayers of topological insulators and ferromagnets using ST-FMR and second harmonic techniques. 
The spin Hall angle values from first-principles calculations tend to underestimate the measured values. This is because in experiments, mechanisms other than the intrinsic spin Hall effect may be present, which are not included in our theoretical calculations. Yet, theoretical results are within an order of magnitude of measured values.
Overall, the first-principles estimates of the spin Hall angle suggest that the intrinsic bulk contribution plays a significant role in spin generation and magnetization switching in bilayers of topological insulators and thin-film magnets. 
We acknowledge the limitation of first-principles study in terms of accuracy. 
The general band gap problem of density-functional theory calculations, certainly affects the calculations but the effect on the energy levels below the Fermi energy is minimal. 
Another issue pertains to the accuracy of the Brillouin zone integration. 
Since these materials have a large unit cell and, therefore, a large Wannier basis, the integration over the Brillouin zone is quite demanding. 
We have utilized an adaptive integration scheme to achieve an optimal tradeoff between complexity and accuracy. In our calculations, the numerical error is estimated to be less than 10\% for a reasonable mesh size on a large CPU cluster (see Supplemental Material for details of the spin Hall conductivity calculations). For a higher accuracy, one might need additional computational power. 

\section*{Acknowledgements}
This work was supported in part by the Semiconductor Research Corporation (SRC) and the National Science Foundation (NSF) through Grant No. ECCS 1740136.
S.M. Farzaneh would like to thank Soheil Abbasloo at New York University
for his generosity in spending time on helping with the software setup and debugging codes.

\appendix*
\section{\label{sec:app}Computational Details}
First-principles calculations are performed within the framework of density functional theory which is implemented in Quantum ESPRESSO suite~\cite{giannozzi_quantum_2009, giannozzi_advanced_2017}. 
Projector augmented-wave\cite{blochl_projector_1994} pseudopotentials~\cite{dal_corso_pseudopotentials_2014} are utilized to reduce the cutoff energies and improve computational efficiency. 
The detailed setup description and parameters are listed in Table \ref{tb:abinitio-setup}. 
The crystal parameters of Sb$_2$Se$_3$ are obtained from Ref. \cite{cao_rhombohedral_2018}. The initial crystal parameters and atomic positions of the other three compounds are obtained from Materials Project \cite{jain_commentary_2013}. The structural relaxation is preformed on the crystals to set the total force to zero. The crystal parameters such as the lattice constant $a$ and the angle between primitive vectors $\alpha$ along with the relaxed atomic positions in the unit cell are provided in Table \ref{tb:abinitio-coordinates}. 

\bgroup
\def\arraystretch{1.5}
\begin{table}[ht]
\centering
\caption{Setup parameters of the first-principles calculations and post processing Wannier methods.}
\label{tb:abinitio-setup}
\begin{tabular}{p{1.05in}cccc}
\hline
& Bi$_2$Se$_3$ & Bi$_2$Te$_3$ & Sb$_2$Se$_3$ & Sb$_2$Te$_3$ \\
\hline
Pseudopotential Type & \multicolumn{4}{c}{Projector Augmented Waves\cite{blochl_projector_1994}} \\
Exchange-Correlation functional & \multicolumn{4}{c}{Generalized gradient approximation\cite{perdew_generalized_1996}}\\
Kinetic $E_\text{cut}$ (Ry)& 56 & 56 & 55 & 34\\
Charge $E_\text{cut}$ (Ry)& 457 & 457 & 249 & 242\\
$k$ mesh & \multicolumn{4}{c}{$8\times8\times8$}\\
Number of Bands & 90 & 90 & 70 & 70\\
\hline
Wannier projections & \multicolumn{4}{c}{$s$ and $p$ orbitals} \\
Wannier $k$ mesh & \multicolumn{4}{c}{$50\times50\times50$ (adaptive $4\times4\times4$ mesh)} \\
\hline
\end{tabular} 
\end{table}

\begin{table}[ht]
\centering
\caption{The parameters describing the crystal structure including the lattice constant $a$, the angle between primitive vectors $\alpha$, and the $z$-component of the atomic positions of the five atoms in the primitive unit cell. The positions are relative to the most bottom atom denoted in Fig. \ref{fig:crystal} and are in Cartesian coordinates (in \AA~units).}
\label{tb:abinitio-coordinates}
\begin{tabular}{cccccccc}
\hline
& $a$ (\AA) & $\alpha$ ($^\circ$) & $z_1$ & $z_2$ & $z_3$ & $z_4$ & $z_5$ \\
\hline
Bi$_2$Se$_3$ & 10.27 & 23.56 & 0.00 & 6.44 & 11.92 & 18.01 & 23.49\\
Bi$_2$Te$_3$ & 10.64 & 24.20 & 0.00 & 6.53 & 12.37 & 18.59 & 24.43\\
Sb$_2$Se$_3$ & 10.01 & 23.53 & 0.00 & 6.32 & 11.60 & 17.59 & 22.87 \\
Sb$_2$Te$_3$ & 10.74 & 23.26 & 0.00 & 6.76 & 12.44 & 18.91 & 24.59 \\
\hline
\end{tabular} 
\end{table}
\egroup

The Bloch basis of the first-principles results are converted to the Wannier basis by projecting into $s$ and $p$ orbitals, which comprise the bands in the vicinity of the Fermi level.
The initial projected Wannier functions are optimized to obtain a maximally localized set via Wannier90 code~\cite{pizzi_wannier90_2020}. Utilizing the post processing module of the Wannier90 code developed by Ref. \cite{qiao_calculation_2018}, the maximally localized Wannier functions are then used to calculate the spin Hall conductivity by evaluating the matrix elements that appear in the Kubo formula and integrating the Berry-like curvature over the Brillouin zone. 
The numerical details of the Wannier methods are listed in Table \ref{tb:abinitio-setup} as well. 

\bibliography{_manuscript-ti-spin-hall}

\pagebreak
\clearpage
\onecolumngrid

\setcounter{section}{0}
\setcounter{equation}{0}
\setcounter{figure}{0}
\setcounter{table}{0}
\renewcommand{\thesection}{S\arabic{section}}
\renewcommand{\theequation}{\thesection.\arabic{equation}}
\renewcommand{\thefigure}{S\arabic{figure}}
\renewcommand{\thetable}{S\arabic{table}}
\section*{Supplementary Information}
In this document, we first review the linear response theory and provide a brief derivation of the Kubo formula for the spin Hall conductivity. Computational details such as the computation time for different stages of the calculations are provided along with a discussion of the accuracy of the calculations and the number of $k$ points in the mesh required for Wannier interpolation. 
\section{Linear response theory and the Kubo formula}
In the linear response theory the response of a system to a perturbation is assumed to be dominated by a linear function of the perturbation which is generally a tensor called linear response tensor. 
Whenever the perturbation is an electric field and the system responds by generating a spin current, the tensor is called spin conductivity and its off-diagonal components represent the spin Hall conductivity. 
For a general Hamiltonian $H=H_0 + H'(t)$ decomposed into an unperturbed time-independent term $H_0$ and a time-dependent perturbation $H'(t)$, the Kubo formula goes as follows. The change in the expectation value of a given observable $\mathcal{O}(t)$, to the linear order in $H'(t)$, can be written in terms of a correlation function, that is
\begin{equation}
    \label{eqs:correlation}
    \delta\ev{\hat{\mathcal{O}}(t)} = -\frac{i}{\hbar}\int_{-\infty}^{\infty} dt'\theta(t - t') \ev{\comm{\hat{\mathcal{O}}(t)}{H'(t')}}_0,
\end{equation}
where the observable $\hat{\mathcal{O}}(t)$ is in the interaction picture defined as $e^{iH_0t/\hbar}\mathcal{O}e^{-iH_0t/\hbar}$. The correlation function includes a $\ev{}_0$ which denotes an ensemble average over the occupied states of the unperturbed $H_0$. 
Taking the Fourier transfer of the above equation results in
\begin{equation}
    \label{eq:correlation-fourier}
    \delta\ev{\hat{\mathcal{O}}_\omega} = -\frac{i}{\hbar}\int_{0}^{\infty} dte^{i\omega t}\ev{\comm{\hat{\mathcal{O}}(t)}{H'_\omega}}_0\cdot 
\end{equation}
Replacing $\hat{\mathcal{O}}(t)$ with $e^{iH_0t/\hbar}\mathcal{O}e^{-iH_0t/\hbar}$ and inserting a completeness relation $\sum\op{m}{m}=\mathbb{I}$, one obtains 
\begin{equation}
    \label{eqs:correlation-completeness}
    \delta\ev{\hat{\mathcal{O}}_\omega} = -\frac{i}{\hbar}\int_{0}^{\infty} dte^{i\omega t}\sum_{n,m}f(\epsilon_n)\qty[\mel{n}{\mathcal{O}}{m}\mel{m}{H'_\omega}{n}e^{i(\epsilon_n - \epsilon_m)t/\hbar} -  \mel{n}{H'_\omega}{m}\mel{m}{\mathcal{O}}{n}e^{i(\epsilon_m - \epsilon_n)t/\hbar}],
\end{equation}
where $f(\epsilon_n)$ is the Fermi-Dirac distribution function. Performing the integral results in
\begin{equation}
    \label{eqs:correlation-completeness}
    \delta\ev{\hat{\mathcal{O}}_\omega} = -i\sum_{n,m\not= n}f(\epsilon_n)\qty[\frac{\mel{n}{\mathcal{O}}{m}\mel{m}{H'_\omega}{n}}{\epsilon_n - \epsilon_m + \hbar\omega} -  \frac{\mel{n}{H'_\omega}{m}\mel{m}{\mathcal{O}}{n}}{\epsilon_m - \epsilon_n + \hbar\omega}],
\end{equation}

\section{Kubo formula for spin Hall conductivity}
In the spin Hall effect, the response of the system is a spin current operator defined as $\mathcal{O} = \mathcal{J}_\alpha^\gamma = (\hbar/2)\{v_\alpha,\sigma_\gamma\}/2$. 
A time varying electric field represented by its frequency components $E_\beta e^{-i\omega t}$ results in $H'_\omega = -\vb*{J}\cdot\vb*{A}$ where $\vb*{J}=-ev_\beta$ is the charge current operator and $\vb*{A} = E_\beta e^{-i\omega t}/i\omega$ is the vector potential. Therefore, by replacing  $H'_\omega$ with $E_\beta/i\omega$ and expanding the $(\epsilon_n - \epsilon_m + \hbar\omega)^{-1}$ terms at the $\omega\rightarrow 0$ limit, one obtains the spin Hall conductivity $\sigma_{\alpha\beta}^\gamma = \delta\ev{\mathcal{J}_\alpha^\gamma}/E_\beta$ as follows 
\begin{equation}
    \label{eqs:spin-hall-tensor}
    \sigma_{\alpha\beta}^\gamma = i\hbar(-e\hbar/2)\sum_{n,m}f(\epsilon_n)\qty[\frac{\mel{n}{\{v_\alpha,\sigma_\gamma\}/2}{m}\mel{m}{v_\beta}{n} - \mel{n}{v_\beta}{m}\mel{m}{\{v_\alpha,\sigma_\gamma\}/2}{n}}{(\epsilon_n - \epsilon_m)^2}],
\end{equation}
Relabeling the energy eigenstates with the Bloch functions $\ket{n,\vb*{k}}$ for a crystalline system and rearranging the sums one obtains
\begin{equation}
    \label{eqs:spin-hall-bloch}
    \sigma_{\alpha\beta}^\gamma =
    -\qty(\frac{e^2} {\hbar})\qty(\frac{\hbar}{2e})\int\frac{d^3\vb*{k}}{(2\pi)^3}\sum_{n} f(\epsilon_{n,\vb*{k}})\Omega_{\alpha\beta,n}^\gamma(\vb*{k}),
\end{equation}
where $\Omega_{\alpha\beta,n}^\gamma(\vb*{k})$ is called the \textit{spin} Berry curvature where 
\begin{equation}
\label{eqs:spin-hall-berry}
    \Omega_{\alpha\beta,n}^\gamma(\vb*{k}) = \hbar^2\sum_{m\not=n} \frac{-2\Im{\mel{n\vb*{k}}{\acomm{v_\alpha}{\sigma_\gamma}/2}{m\vb*{k}} \mel{m\vb*{k}}{v_{\beta}}{n\vb*{k}}}}{(\epsilon_{n,\vb*{k}} - \epsilon_{m,\vb*{k}})^2}\cdot
\end{equation}
We note that both the spin and the charge current operators need to be divided by the volume of the system to give the correct units of the current density. However, since there are two sums over the Brillouin zone, the volumes would be cancelled as the sums are rewritten in terms of $k$-space integrals.

\section{Computational details}
The first principles calculations of the common topological insulators with the formula A$_2$B$_3$ are quite demanding because of their more complex crystal structure compared to that of heavy metals such as Pt. Also, since there are five atoms in the unit cell of these compounds, compared to one atom in Pt, the volume of the unit cell is about an order of magnitude larger than that of Pt. 
Therefore, one expects to see about an order of magnitude longer computation times compared to the case of Pt. 
Here we report the computation time for different stages of the first principles calculations such as the structural relaxation (to achieve zero total force), the self-consistent and non-self-consistent field calculations (to obtain eigenstates on a regular basis), maximally localized Wannier set by iterative optimization, and finally integrating the Berry-like curvature of the spin Hall conductivity over the entire Brillouin zone. The computation times are listed in Table \ref{tb:time} for all the four topological insulators considered in this work. We note that these reported times are for a cluster of 96 CPUs. 
As seen from the table above the last stage, which involves the integral over the Brillouin zone, is computationally the most demanding one.  
\bgroup
\def\arraystretch{1.5}
\begin{table}[ht]
\centering
\caption{Computation time of different stages of the first principles calculations and the evaluation of spin Hall conductivity in the Wannier basis. The values are in seconds. These units are obtained on a cluster of 96 CPUs.}
\label{tb:time}
\begin{tabular}{p{3.0in}cccc}
\hline
& Bi$_2$Se$_3$ & Bi$_2$Te$_3$ & Sb$_2$Se$_3$ & Sb$_2$Te$_3$ \\
\hline
Structural Relaxation & 1095 s & 90 s & 870 s & 640 s \\
Self consistent field (Monkhorst $8\times8\times8$ mesh) & 380 s & 780 s & 960 s & 720 s \\
Non-self consistent field (regular $8\times8\times8$ mesh) & 1320 s & 2670 s & 3600 s & 2365 s \\
Maximally localized Wannier set (\# of iterations) & 9400 s (32000) & 11000 s (32000) & 4000 s (12500) & 10660 s (32000)\\
Wannier $k$ mesh ($50\times50\times50$ (adaptive $4\times4\times4$ mesh)) & 70850 s & 107235 s & 27850 s & 44040 s \\
\hline
\end{tabular} 
\end{table}
\egroup

\section{Choosing the mesh size}

\begin{figure}
    \begin{tabular}{p{2.23in}p{2.23in}}
    \includegraphics[]{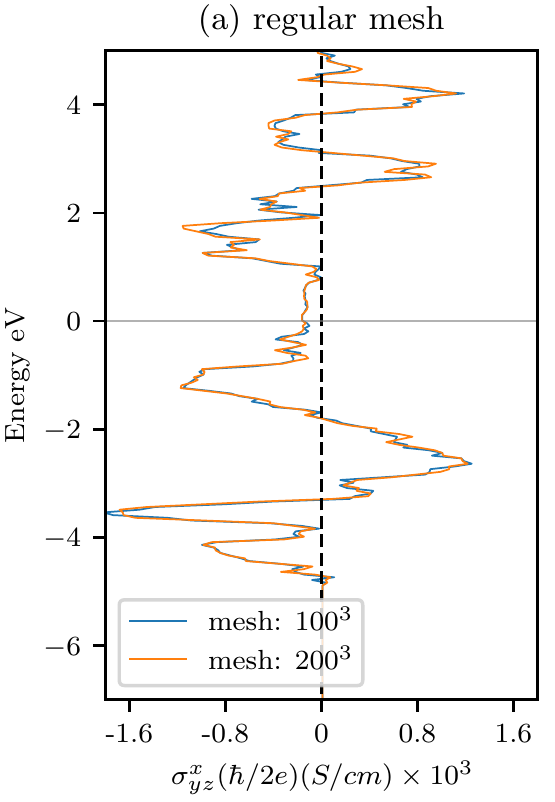} & \includegraphics[]{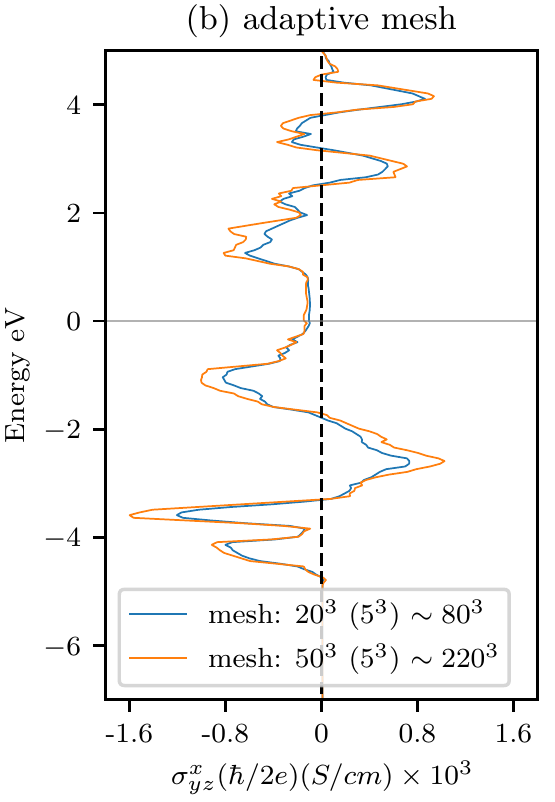} 
    \end{tabular}
    \caption[example] 
    {The effect of the size and the shape of the integration mesh comparing  the regular (a) and the adaptive (b) meshes for Bi$_2$Se$_3$. The adaptive mesh leads to a smoother result and a faster convergence than the one obtained by the regular mesh due to the more efficient distribution of the k points in the vicinity of the spikes.}
    \label{fig:mesh} 
\end{figure}  

\begin{figure}
    \includegraphics[]{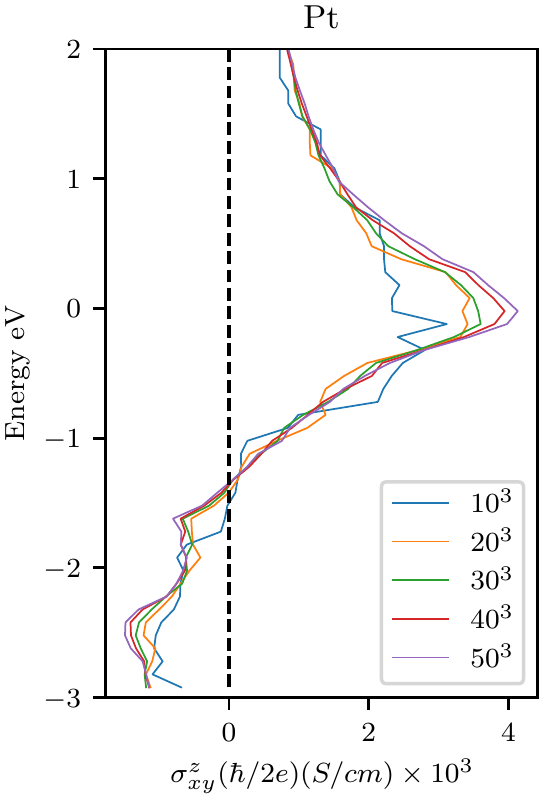} 
    \caption[example] 
    {The effect of the mesh size on the spin Hall conductivity of Pt. The legend shows the initial size of the adaptive mesh with a $5\times5\times5$ additional mesh around the spikes. The finest mesh reproduces the previous results reported in Refs. \cite{guo_intrinsic_2008} and \cite{qiao_calculation_2018} within a good accuracy (within a 5\% difference).}
    \label{fig:mesh-pt} 
\end{figure}  

It has been shown elsewhere \cite{qiao_calculation_2018, guo_intrinsic_2008} that the number of the $k$ points required for a fairly accurate integration over the Brillouin zone is of the order of $10^6$. 
Fig. \ref{fig:mesh}a shows $\sigma_{zy}^x$ for Bi$_2$Se$_3$ on a regular mesh with $100\times100\times100$ and $200\times200\times200$ points, respectively. As seen from the figure the results are very noisy. The reason is that the kernel of the Brillouin zone integral contains several spikes due to accidental degeneracies in the band structure. Therefore, a very fine mesh is needed to capture all those spikes. To alleviate this problem one could resort to an adaptive way of choosing the k points of the mesh. In the adaptive method, the mesh gets finer in the vicinity of a spike while it is coarser in the smoother areas. This way, the spikes can be taken into account more accurately. Figure \ref{fig:mesh}b shows the result for the adaptive mesh which converges for a mesh of $50\times50\times50$ (with an additional $5\times5\times5$ mesh around each spike) and is much smoother than the one obtained from a regular mesh with fixed distancing, for an approximately equal number of points in total. 
For example, a mesh of $50\times50\times50$ with adaptive meshing of $5\times5\times5$ around the spikes contains about $10^7$ total number of points which is of the same order of magnitude as that of a $200\times200\times200$ regular mesh but captures the spike better than the regular mesh. This suggests that with roughly equal number of points, one can achieve a more accurate result by distributing the $k$ points more efficiently using the adaptive meshing method. 
Furthermore, we reproduce the results previously reported for platinum by comparing meshes with different sizes in Fig. \ref{fig:mesh-pt}. As the mesh gets finer, the values of the spin Hall conductivity converge at each energy level. This figure shows that an adaptive mesh with $50^3$ initial points and $5^3$ additional points at each spike can reproduce a good match (within a 5\% error) with the results reported in Refs. \cite{guo_intrinsic_2008} and \cite{qiao_calculation_2018}. 

\section{Convergence and error}
\begin{figure}
    \includegraphics[]{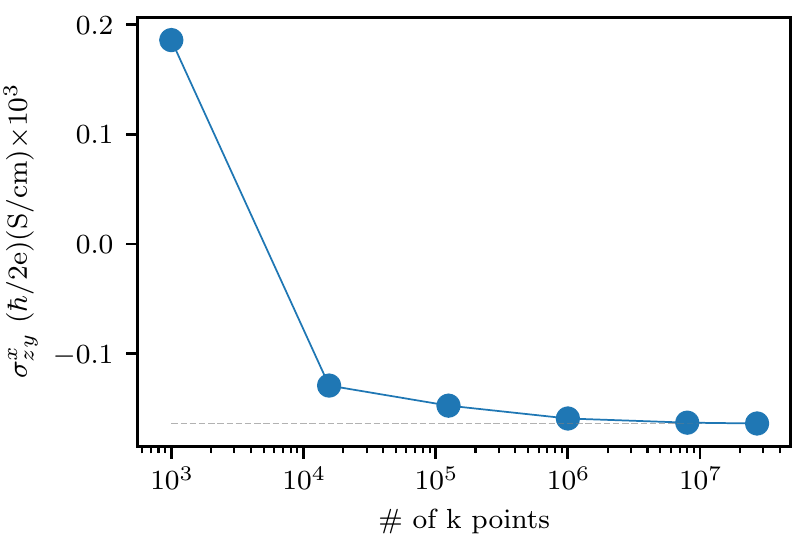}  
    \caption[example] 
    {The convergence of the spin Hall conductivity of Bi$_2$Se$_3$ at the Fermi energy for different number of k points on the mesh. Assuming that the value for a mesh of $300^3$ points is the converged value, that is  $-1.63\times10^2$, the relative error for the mesh with size $50^3$ is -9.8\%.}
    \label{fig:convergence} 
\end{figure}  

In order to make sure that the finite value of spin Hall conductivity at the Fermi energy is not a numerical error, we perform convergence tests in which we observe the behavior of $\sigma_{zy}^x$ as the mesh size increases. Figure \ref{fig:convergence} plots $\sigma_{zy}^x$ versus the number of k points in the mesh considering the mesh is regular and not adaptive. 
If the we consider the last value corresponding to the finest mesh as the converged value $\sigma_{zy}^x=-1.63\times10^2$, the relative error for the $50\times50\times50$ is -9.8\%.

\end{document}